\documentstyle[11pt,mrs2003,epsfig]{article}

\newcommand{\lsim}{\,\lower2truept\hbox{${<\atop\hbox{\raise4truept\hbox{$
\sim$}}}$}\,}
\newcommand{\gsim}{\,\lower2truept\hbox{${>\atop\hbox{\raise4truept\hbox{$
\sim$}}}$}\,}

\begin{document} 
 
\title{WEAK LENSING, STRUCTURE FORMATION AND DARK ENERGY}

\author{Fabio Giovi, Carlo Baccigalupi and Francesca Perrotta}
\affil{SISSA/ISAS Via Beirut 4, 34014, Trieste, Italy}

\begin{abstract}
We study how CMB bispectrum, produced by weak gravitational lensing
and structure formation, can constrain the redshift evolution of the
dark energy equation of state independently on its present value. 
Analyzing the line of sight contribution to the angular CMB
bispectrum, we find that the relevant redshift at which the structure 
formation contributes to the signal is $0.1 \lsim z \lsim 2$ for
multipoles $1000 \gsim l \gsim 100$: just the epoch when the dark 
energy starts to dominate the cosmological expansion rate. For
scenarios having the same equation of state at the present, this turns
out to be a new observable capable to discriminate between models of
dark energy with different time evolution of equation of state. We
assess the strength of this effetc within the framework of tracking 
Quintessence trajectories obeying SUGRA and Ratra-Peebles potentials.
\end{abstract}
 
\section{Introduction}
In the last few years the cosmological parameters have been constrained 
in a narrow range of values by several experiments and the picture of 
the universe seems to be the following. The unexpected dimming of type
Ia supernovae (SNIa), used as standard candles, is the evidence of an 
accelerating expansion of the universe, SNIa probes the cosmological 
expansion up to redshift $z \sim 2$ \cite{RIESS,PERLM}; in a Friedmann 
Robertson Walker universe this acceleration is provided by an amount
of dark energy that is $70\%$ of the whole energy content. Other
independent indications of the presence of a missing energy come from 
Cosmic Microwave Background (CMB) anisotropies \cite{SPERG} and Large 
Scale Structures (LSS) \cite{DODEL}, CMB data support a flat
cosmological geometry while LSS indicate that $30\%$ of critical
density is clustered. The first candidate for the dark energy is the 
Cosmological Constant, which has two well known  problems: the
coincidence problem (why Cosmological Constant density and matter
density are comparable today?) and the fine tuning problem (why
Cosmological Constant scale is 120 orders of magnitude less than
Planck scale?). To solve the latter problem a dynamical scalar field,
known as Quintessence \cite{STEIN,PEEBL}, has been introduced as a
minimally extension of Cosmological Constant. With this field we have
a dynamical equation of state for the dark energy alleviating the fine 
tuning problem. Recently a strong constraint has been put for the
value of equation of state at present, combining CMB, LSS, SNIa and
HST observations we have $w_{0}<-0.78$ \cite{SPERG}. The next
challenge in cosmology is to constrain the time evolution of $w$; this
can be done with future experiments like PLANCK for CMB \cite{BALBI}
and future SNIa observations.

The weak lensing effect on CMB anisotropies has been studied \cite{KOMA}
and the third order statistics, the bispectrum, has been used to
constraint the effective dark energy equation of state \cite{VERDE}. If the 
primordial anisotropies are Gaussian the bispectrum is zero within 
cosmic variance; cosmic structures produce a non-Gaussian feature on 
CMB and a non vanishing bispectrum. In this work we study the weak
lensing effect produced by structure formation on CMB pattern in
Quintessence cosmology; the bispectrum is written in terms of a line
of sight integral involving the redshift derivative of gravitational
potential and cosmological distances. Since the dynamics of dark
energy involves both perturbations growth and cosmological distances,
also the bispectrum depends on the evolution of dark energy; here we
show how different dark energies with same value of $w$ at present
produce different features on the bispectrum, introducing a new
observable capable to discriminate between different dark energies 
independently on the value of $w_{0}$ \cite{GIOVI}.

\section{Quintessence, weak lensing and integrated Sachs-Wolfe effect}
We have considered two kind of potential for Quintessence: the
Ratra-Peebles inverse power law (hereafter RP \cite{RP}) and SUGRA 
\cite{SUGRA}. In terms of scalar field $\phi$ the potentials are written as 
\begin{equation}
\label{e:potentials}
V_{{\it RP}}=\frac{M^{\left( 4+\alpha \right)}}{\phi^{\alpha}},
\ \ V_{{\it SUGRA}}=\frac{M^{\left( 4+\alpha \right)}}
{\phi^{\alpha}}\cdot 
e^{4\pi G\phi^{2}}\ ;
\end{equation}
with these potentials the problem of fine tuning is reduced: the only 
relevant condition for the initial value of scalar field is 
$\phi_{i}<<M_{Pl}$ ($M_{Pl}$ is the Planck mass). In this way we can
reach the present value $\phi_{0}$ for a wide set of initial condition 
\cite{PEEBL}; in the case of a scalar field the equation of state
$w_{Q}$, defined as $p_{Q} / \rho_{Q}$, evolves in time
depending on the potentials.

We consider now the CMB anisotropies produced at decoupling and those 
produced by the integrated Sachs-Wolfe effect (hereafter ISW
\cite{ISW}) taking into account the weak lensing effect of growing
structures. The CMB anisotropy in the direction $\hat{n}$ can be
decomposed as
\begin{equation} \label{e:dt}
\Theta \left( \hat{n} \right) \simeq \Theta_{lss} \left( \hat{n} 
\right) + \Theta_{ISW} \left( \hat{n} \right) + \vec{\nabla} 
\Theta_{lss} \left( \hat{n} \right) \cdot \vec{\alpha}\ , 
\end{equation}
the first term is the primordial anisotropy, the second one is the
ISW anisotropy and the last one is the contribution due to weak
lensing; $\alpha$ is the deflection angle. The potential perturbations 
that produces the ISW produces also the weak lensing; because of this the
lensed CMB photons and those redden by ISW are correlated and a
non-Gaussian feature on CMB pattern is produced, that give us a non
null signal of bispectrum. To evaluate the non linear density power 
spectrum we have used a semi-analytical approach \cite{MA}. Throughout
this work we used the best fit of cosmological parameters (excluding
$w$) \cite{BENNE}, but with a $n_{s}=1$ value for the spectral index.

\section{The CMB bispectrum}
To calculate the bispectrum, we have used three equal multipoles $l$
probing the CMB pattern with equilateral triangles with size of about 
$\theta \sim 180^{o} / l$; this is the easiest choice but it is 
sufficient to show how the dynamics of dark energy affect the
bispectrum. The CMB bispectrum is written as
\begin{equation} \label{e:bispectrum}
B_{l}=3l \left( l+1 \right) \sqrt{\frac{\left( 2l+1\right)^{3}}{4\pi}} 
\left(
\begin{array}{ccc}
l & l & l \\
0 & 0 & 0
\end{array}
\right) 
C_{l}^{P} Q \left( l \right)\ ,
\end{equation}
where $C_{l}^{P}$ is the primordial power spectrum at multipole $l$, the
parenthesis are the Wigner's 3J symbols and $Q\left(l\right)$ is
\begin{equation} \label{e:ql}
Q \left( l \right) \equiv \left< \left( a_{lm}^{lens} \right)^{*}
a_{lm}^{ISW} \right> \simeq 2 \int_{0}^{z_{lss}} dz \frac{r \left(
  z_{lss} \right) - r \left( z \right)}{r \left( z_{lss} \right)r^{3}
  \left( z \right)} \left[ \frac{\partial P_{\Psi} \left( k,z
    \right)}{\partial z} \right]_{k=\frac{l}{r\left(z\right)}}.
\end{equation}
In the previous relation the quantity $P_{\Psi}\left(k,z\right)$ is
the gravitational potential power spectrum, related to the density
power spectrum by the relation $P_{\Psi} \left( k,z \right)=\left(
\frac{3}{2} \Omega_{M0} \right)^{2} \left( \frac{H_{0}}{ck}
\right)^{4} P \left( k,z \right) \left( 1+z \right)^{2}$, $z_{lss}$ 
is the redshift of last scattering surface and $a_{lm}^{lens}$ and 
$a_{lm}^{ISW}$ are respectively the coefficients of harmonics
expansion of lensing contribution and ISW. A detailed outline on how
write Eq. (\ref{e:bispectrum}) and (\ref{e:ql}) can be found in
\cite{VERDE} and references therein.

\begin{figure} 
\vspace*{1.25cm}  
\begin{center}
\epsfig{figure=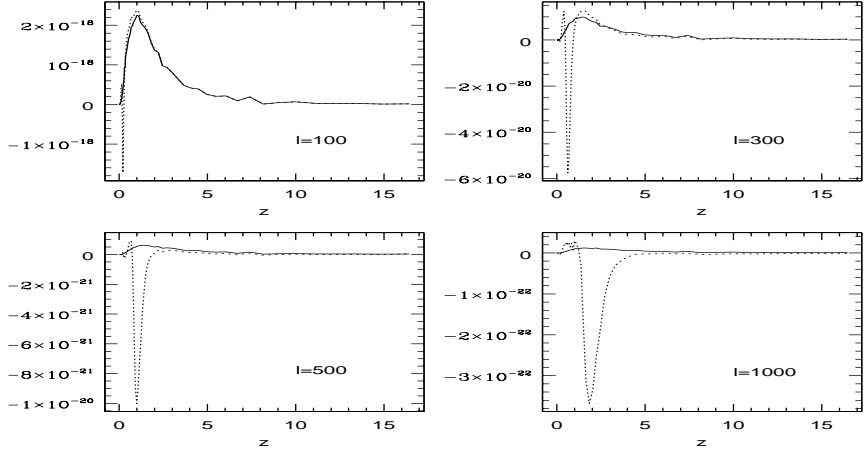,width=12cm,height=6.5cm}  
\end{center}
\vspace*{0.25cm}  
\caption{Integrand of (\ref{e:ql}) for SUGRA model with $w_{0}=-0.9$ for 
different multipoles; solid line is linear regime, dotted line is non
linear regime. The graphics are qualitative the same for constant $w$ and RP.} 
\label{g:panel}
\end{figure}

The integral $Q\left(l\right)$ is the most relevant quantity in this
work, it describes how the weak lensing on CMB is produced by the
forming structures along the line of sight. Analyzing the redshift 
behavior of the integrand of (\ref{e:ql}) we can study how the 
contribution of $Q\left(l\right)$ is distributed along the line of 
sight and how it depends on the time evolution of Quintessence 
equation of state. The integrand can be decomposed in two factors: 
a geometrical factor $\frac{r\left(z_{lss}\right)-r\left(z\right)}
{r\left(z_{lss}\right)r\left(z\right)^{3}}$ and a derivative factor 
$\frac{\partial P_{\Psi}}{\partial z}$. In the geometrical factor the 
redshift behavior of $w$ is averaged through the distance integral so
we don't have a significant sensitivity to this parameter while the derivative
factor is more sensitive to the time evolution of Quintessence
equation of state giving the most important contribution to the
integrand of $Q\left(l\right)$. In Fig. \ref{g:panel} we have plotted
the integrand of (\ref{e:ql}) for SUGRA model with $w_{0}=-0.9$, once we
fix the multipole $l$; coming form high redshift, the integrand probes
from large to small scales by the relation $\lambda\left(z\right)=2\pi
r\left(z\right)/l$. As the redshift comes down from infinity, the
scale $\lambda\left(z\right)$ decreases until it meets the growing 
comoving scale of the horizon becoming a sub-horizon comoving scale;
the gravitational potential starts to decrease because of the free 
streaming and this feature is represented by the positive peaks in
Fig. \ref{g:panel} (the derivative factor is positive). As $z$ get
smaller, the scale $\lambda\left(z\right)$ matches the scales which
enter in the non-linear phase at that epoch and the power increase
with time (the derivative factor is negative); this feature is shown
in Fig. \ref{g:panel} with a negative peak. When $z$ approaches zero, 
the wavenumber tends to infinity and the power vanishes. 
$Q\left(l\right)$ has a non zero contribution, yielded by the
derivative factor, in the narrow range of redshift (between 0.1 and 2)
where the dark energy starts to dominate over the matter; therefore the
integrand can be used as a window function that selects just the
relevant redshift to probe the cosmic acceleration. In this pictures the dark 
energy equation of state is probed independently on its value today 
(the power is zero at the present as remarked above) and only at the
redshift where the cosmological expansion rate is affected by it.
In this way the correlation (\ref{e:ql}) is a powerful tool to study the 
redshift behavior of dark energy equation of state; RP and SUGRA have
a different evolution of equation of state, from the tracking regime
to $w_{0}$, and this difference is just in the range of redshift where
the integrand is not null.

\begin{figure}
\vspace*{1.25cm}  
\begin{center}
\epsfig{figure=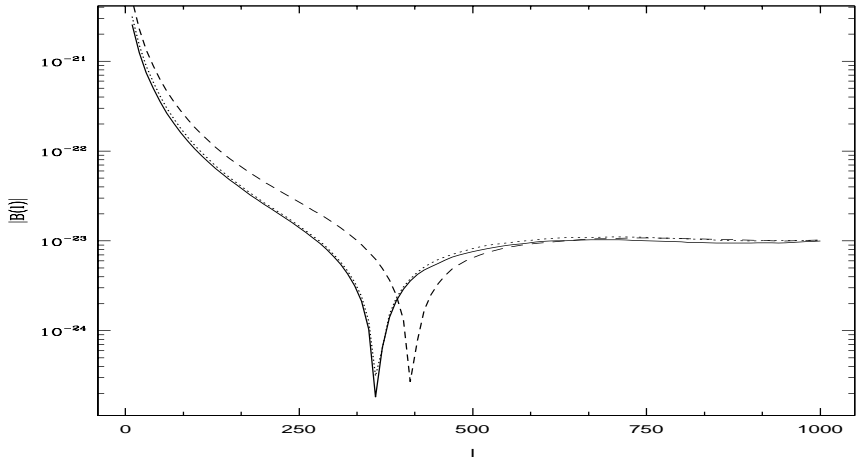,width=12cm,height=6.5cm}  
\end{center}
\vspace*{0.25cm}  
\caption{Absolute value of CMB bispectrum in Mpc$^{-3}$. Solid line 
is constant $w$, dotted line is RP, dashed line is SUGRA; all models 
have the same value of equation of state at present.} 
\label{g:bispectra}
\end{figure}

\begin{figure} 
\vspace*{1.25cm} 
\begin{center}
\epsfig{figure=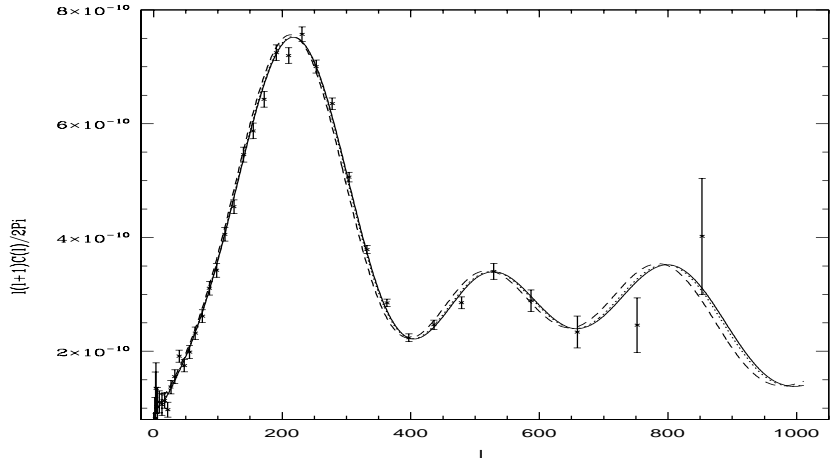,width=12cm,height=6.5cm}  
\end{center}
\vspace*{0.25cm}  
\caption{Power spectra for constant $w$ (solid line), RP (dotted line)
  and SUGRA (dashed line) $w_{0}=-0.9$, error bars are binned WMAP
  data.}
\label{g:spectra}
\end{figure}

The different behaviors of $w\left(z\right)$ for different models of
dark energy are reflected in $Q\left(l\right)$ and also in the
bispectrum. In Fig. \ref{g:bispectra} we plotted the bispectrum for
constant $w$, RP and SUGRA model, all with $w_{0}=-0.9$. The main feature of 
bispectrum is the cusp and represents the transition between linear
and non linear regime when density perturbations grows i.e. when the 
integral (\ref{e:ql}) is zero. The position of cusp in $l$-space
depends on the dark energy model considered; in SUGRA model the cusp
is shifted with respect to RP models of 50 multipoles in spite of the
fact that both models have the same value of $w_{0}$ (for SUGRA the cusp is
located at $l=412$, for RP at $l=362$). This shift is due to the fact
that in SUGRA scenario the growth factor is sensibly larger than that
of RP scenario at almost all epochs \cite{GIOVI}. Correspondingly,
the change at low redshift (when the dark energy starts to dominate
the expansion) is stronger in SUGRA scenario producing a larger
amplitude for the linear regime in the line of sight integral 
(when the derivative factor is positive). The contribution to the
bispectrum coming from the non linear scales has lower amplitude
because the rise due to the onset of non-linearity has to overcome the 
gravitational potential decay which in SUGRA is stronger than in RP. 
The net effect is that, for a given multipole $l$, the larger is the 
value of $w$ at the relevant epoch, the larger is the contribution
from the linear regime; the scale at which linear power balances the 
non-linear one is shifted at larger wavenumber (higher multipoles) as
is shown in Fig. \ref{g:bispectra}. The dependence on the dark energy
of the position of the cusp in the bispectrum is a new feature that
can be used to discriminate between different dark energies and must be
take in consideration in future observations. For completness, in
Fig. \ref{g:spectra} we plot the CMB power spectrum for the models
considered, together with the WMAP data \cite{BENNE}; in fact, comparing Fig. 
\ref{g:bispectra} and Fig. \ref{g:spectra}, the degeneration between 
models is higher in the power spectrum than in the bispectrum \cite{GIOVI}. 

\section{Conclusions} 
We have studied the CMB bispectrum produced by weak lensing and ISW
effect in Quintessence cosmology; we have found a new observable
capable to discriminate between different models of dark energy with
the same value of equation of state at the present but with different 
behavior in the past.

The correlation between weak lensing and ISW is made of two
contributions yielding opposite signs after horizon crossing: in the
linear regime the gravitational potential decreases in time, in the
non-linear one we have a growth. Most of the contribution to the
bispectrum signal come from the narrow range of redshift where the
dark energy starts to dominate the expansion rate of the universe
driving the acceleration and this epoch is that typical of structure 
formation. Because of these reasons, the bispectrum is sensitive to 
cosmic expansion rate at that redshift, i.e. to the equation of state
at that time independently on its value today. To illustrate this
effect we have used two kinds of dark energy: RP model and SUGRA
model. Both RP and SUGRA have a time-variable equation of state but a 
different behavior in redshift. As result we have a geometry shift of the
projected bispectrum, affecting in particular the cusp, that represent
the balance between linear and non-linear regime; in the case of
SUGRA, the cusp is located at higher multipoles with respect RP and
this shift is about one order of magnitude greater than the shift in
the CMB power spectrum.

We don't explored the degeneration of the cusp position with respect
other cosmological parameters, but we must keep in mind that this
effect is due to a projection along the line of sight, so the shift
can be mimicked by variations in the cosmological parameters inducing 
geometrical features in CMB anisotropies (curvature, $H_{0}$). On the 
other hand these variations will strongly affect both the bispectrum
and the power spectrum.

The capability of the next CMB probes, PLANCK and CMBPol, to look for
the effect we pointed out here is currently under study, as well as
future weak lensing surveys \cite{REFRI}.

\acknowledgements{Fabio Giovi would like to thanks Simona Rosati and
  Viviana Acquaviva for encouragement and support.} 


\vfill 
\end{document}